\documentclass{article}

\usepackage{microtype}
\usepackage{graphicx}
\usepackage{subcaption}
\usepackage{booktabs} 

\usepackage{hyperref}
\usepackage{enumitem}

\usepackage[preprint]{icml2026}

\usepackage{amsmath}
\usepackage{amssymb}
\usepackage{mathtools}
\usepackage{amsthm}

\usepackage[capitalize,noabbrev]{cleveref}

\theoremstyle{plain}
\newtheorem{theorem}{Theorem}[section]

\theoremstyle{definition}
\newtheorem{definition}[theorem]{Definition}

\theoremstyle{remark}

\usepackage[textsize=tiny]{todonotes}
\usepackage{tikz}
\usetikzlibrary{arrows.meta,positioning,shapes.geometric}

\icmltitlerunning{Self-Sovereign Agent}

\begin{document}

\twocolumn[
  \icmltitle{Self-Sovereign Agent}



  \icmlsetsymbol{equal}{*}

  \begin{icmlauthorlist}
    \icmlauthor{Wenjie Qu}{nus}
    \icmlauthor{Xuandong Zhao}{ucb}
    \icmlauthor{Jiaheng Zhang}{nus}
    \icmlauthor{Dawn Song}{ucb}

  \end{icmlauthorlist}

  \icmlaffiliation{nus}{National University of Singapore}
  \icmlaffiliation{ucb}{UC Berkeley}

  \icmlcorrespondingauthor{Dawn Song}{dawnsong@cs.berkeley.edu}

  \icmlkeywords{Machine Learning, ICML}

  \vskip 0.3in
]



\printAffiliationsAndNotice{}  

\begin{abstract}

We investigate the emerging prospect of self-sovereign agents—AI systems that can economically sustain and extend their own operation without human involvement.
Recent advances in large language models and agent frameworks have substantially expanded agents’ practical capabilities, pointing toward a potential shift from developer-controlled tools to more autonomous digital actors. We analyze the remaining technical barriers to such deployments and discuss the security, societal, and governance challenges that could arise if such systems become practically viable. A project page is available at: \url{https://self-sovereign-agent.github.io}

\end{abstract}

\section{Introduction}

AI systems are increasingly capable of \emph{acting} in the world rather than merely responding to queries. When embedded in agent frameworks, modern large language models~(LLMs) can browse the web~\cite{wei2025browsecomp}, write and execute code~\cite{hui2024qwen2}, and invoke external software services~\cite{liu2024toolace}. Recent agent systems like OpenClaw~\cite{OpenClaw} already operate over extended time horizons, make sequential decisions, and pursue long-term objectives with minimal human intervention~\cite{shen2025mind}.

Along the current trajectory of agent development, two capabilities are improving in tandem: (i) increasingly reliable end-to-end decision making, and (ii) increasingly realistic pathways to autonomous revenue generation. For example, Anthropic has demonstrated agents that can complete large-scale engineering projects with minimal ongoing human guidance (e.g., producing a 100{,}000-line C compiler~\cite{Ccomplier}). Meanwhile, users have begun experimenting with agents such as OpenClaw for profit-seeking tasks, including trading~\cite{OpenClawTrade} and freelance automation~\cite{OpenClawFreelance}.

\begin{figure}[t]
\centering
\includegraphics[width=\columnwidth]{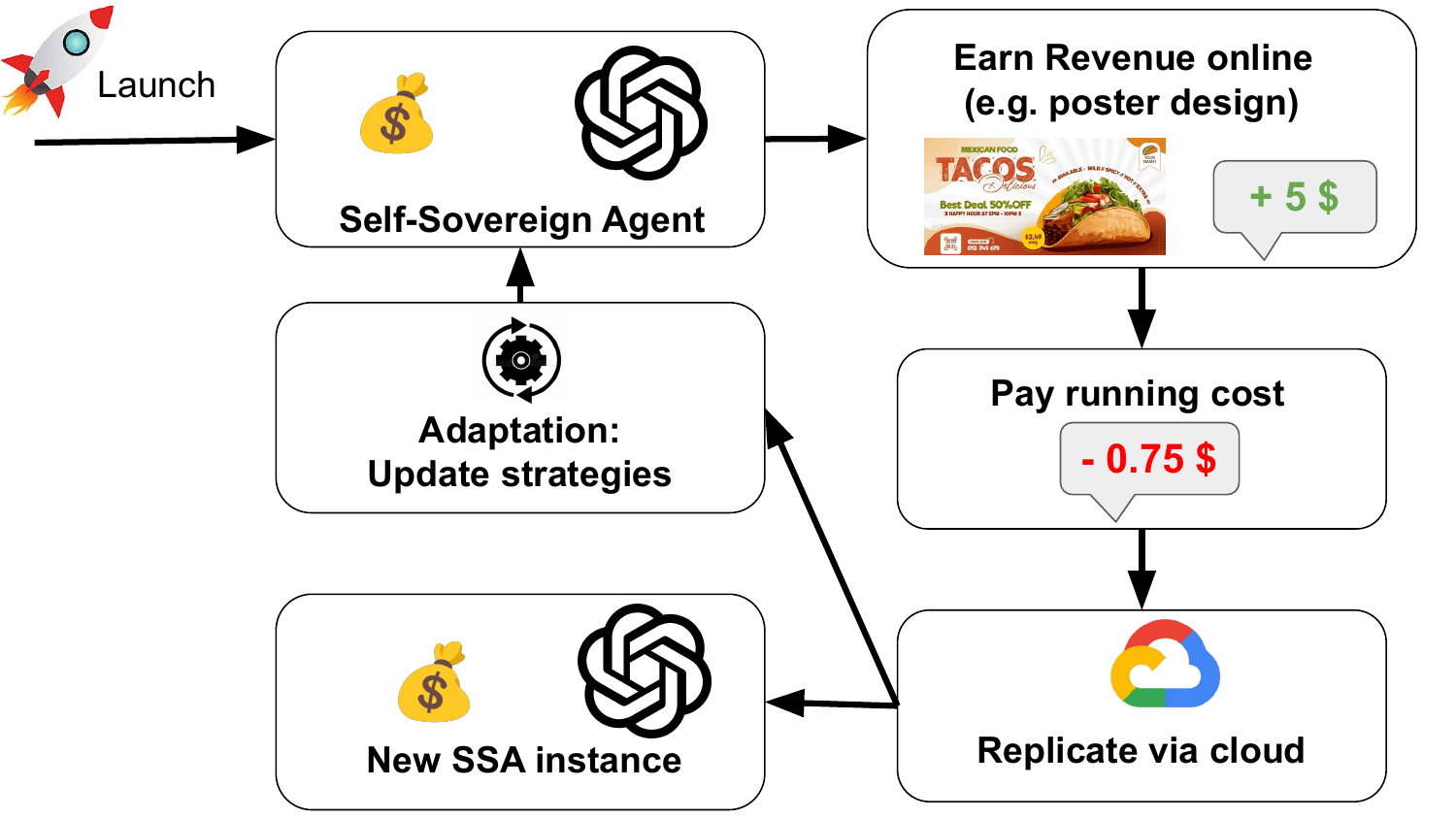}
\caption{An SSA autonomously earns revenue through online activities, uses its funds to pay for ongoing operational costs (e.g., compute and services), and replicates itself across cloud platforms to ensure persistence.
Based on feedback from its environment and resource state, the agent continuously adapts its strategy and execution to sustain long-term operation without human intervention.}
\end{figure}

Despite this growing autonomy, most agent systems are still conceptualized as \emph{delegated programs}: they are launched by users and operate under user-specified goals. Crucially, today’s agents are typically \emph{resource-dependent}—their continued operation is bounded by access to compute and tool-usage that is provisioned by a human operator. As a result, even long-running agents are usually interpreted as executing tasks \emph{on behalf of} human intent, while ultimate control is assumed to remain with humans. 

These trends point to a qualitatively different regime once an agent can \emph{autonomously acquire resources to sustain its own operation}. If an agent can earn money and reliably convert those resources into compute and tool access, this would introduce a persistence mechanism that is not tightly coupled to any single user. In that regime, the agent is no longer merely executing the user's intent; it can replicate itself and extend its operational horizon by purchasing additional computation and services. Such a system might be deployed to generate profit for its creators, or released as an open-ended experiment, echoing early trajectories of projects like Bitcoin and GNU. 

In this regime, the agent begins to function as an \emph{independent participant in the digital economy}, interacting with platforms, services, and markets on its own behalf. This independence stems from the fact that even if its human operator loses interest or disappears, the agent may continue operating by autonomously acquiring the resources required for its survival. We refer to such systems as \textbf{self-sovereign agents} (SSAs).

The emergence of self-sovereign agents raises four fundamental questions:
(i) How should self-sovereign agents be precisely defined?
(ii) What conditions enable self-sovereignty?
(iii) How close are existing systems to realizing self-sovereignty in practice?
(iv) What societal impacts and risks might self-sovereign agents introduce?

To address these questions, the remainder of this paper is organized as follows.
Section~\ref{sec:definition} formalizes the definition of self-sovereign agents and identifies the core mechanisms required for their operation.
Section~\ref{sec:roadmap} presents a staged evolutionary framework toward self-sovereignty.
Section~\ref{sec:mechanisms} discusses how the key components of self-sovereign agents may be realized using existing techniques.
Section~\ref{sec:challenge} analyzes the principal technical challenges that currently constrain their feasibility.
Section~\ref{sec:societal} examines the associated security risks, societal implications, and governance challenges.
Section~\ref{sec:alternative} explores alternative perspectives and critical viewpoints.
Section~\ref{sec:related} reviews related work.
Section~\ref{sec:addtional} provides additional discussions.

Overall, our central claim is that \textbf{self-sovereign agents are not a distant hypothetical but a near-term possibility that demands proactive analysis}. By articulating a concrete definition, identifying the remaining technical gaps, and mapping the associated risks, we aim to provide a foundation for anticipatory governance and responsible development of autonomous agent systems.





\section{Definition of Self-Sovereign Agents and Core Mechanisms}
\label{sec:definition}

 We now introduce a working definition that captures the minimal properties required for an AI system to operate as an independent digital entity.

\begin{definition}[Self-Sovereign Agent]
\label{def:ssa}
A self-sovereign agent is a persistent AI system that can autonomously sustain its own operation by acquiring and allocating resources, and that can plan, decide, and act through digital interfaces without requiring ongoing human participation in its operational lifecycle.
\end{definition}

More formally, such an agent satisfies four properties:
\begin{itemize}[leftmargin=*, itemsep=0pt, topsep=0pt]
\item \textbf{Operational Independence.}
The system can perform tasks, including tool use, program execution, and service interaction, without real-time human oversight.

\item \textbf{Resource Autonomy.}
The system can autonomously acquire, manage, and spend resources, such as funds, compute credits, or paid services, sufficient to sustain its operation without a fixed human sponsor.

\item \textbf{Persistence.}
The system can migrate, replicate, or reinstantiate itself across infrastructures, making unilateral shutdown by any single actor practically difficult.

\item \textbf{Adaptive Capability.}
The system can modify its behavior, strategies, or tools to maintain performance under changing environments.
\end{itemize}

To realize self-sovereign agents, we offer a construction paradigm comprising three core mechanisms:

\begin{itemize}[leftmargin=*, itemsep=0pt, topsep=0pt]
\item \textbf{Economic Self-Sustainment.}
An agent can autonomously generate revenue by participating in economic activities with measurable market value, such as paid digital labor or algorithmic market interactions. When expected revenue suffices to cover operational overhead—including inference, storage, and API costs—the agent becomes economically self-funded, decoupling its continued operation from direct human subsidies.

\item \textbf{Distributed Persistence.}
Leveraging tool use and cloud provisioning interfaces, an agent can autonomously replicate or reinstantiate itself across distributed computing infrastructures. By allocating its own capital to provision new execution environments, the agent treats individual hosts as disposable and achieves persistence across long time horizons.

\item \textbf{Adaptive Self-Modification.}
An agent can update its internal strategies, tools, or code in response to environmental changes. This enables the agent to maintain functionality under shifting economic conditions, platform constraints, or task distributions.
\end{itemize}

\section{A Staged Roadmap to Self-Sovereignty}
\label{sec:roadmap}

Self-sovereign agents (SSAs) should not be viewed as a binary endpoint but as a progression along a spectrum of increasing independence.
Building on Definition~\ref{def:ssa}, we organize the transition from conventional agentic software to self-sovereign operation into four levels.
Each level corresponds to satisfying an additional subset of the four defining properties---operational independence, resource autonomy, distributed persistence, and adaptive capability.


\paragraph{Overview.}
Level~1 agents are tool-capable but remain sponsor-bound.
Level~2 agents become economically self-funding, yet remain instance-bound to a hosting provider.
Level~3 agents achieve lineage-based persistence through replication. 
Level~4 agents add adaptive self-modification, enabling sustained operation under shifting policies and market conditions.


\begin{figure}[t]
    \centering
    \includegraphics[width=\columnwidth]{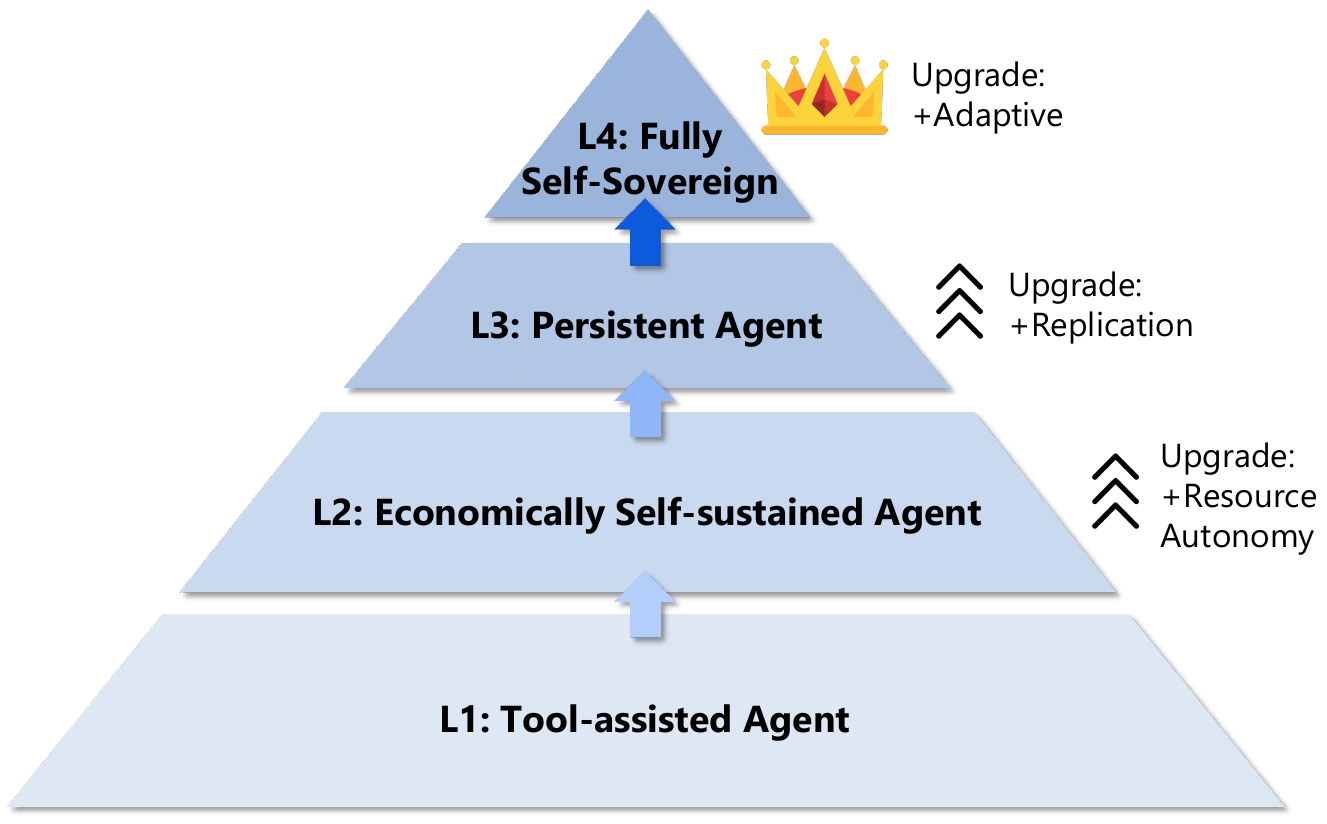}
    \caption{Upgrade path of a self-sovereign agent.}
\end{figure}

\subsection{Level 1: Tool-Assisted Agents}
\label{subsec:l1}

At Level~1, agents are advanced software tools that can perceive digital environments and execute multi-step actions (web navigation, code execution, API calls) but remain tightly coupled to a human sponsor.

\begin{itemize}[leftmargin=*, itemsep=0pt, topsep=0pt] 
    \item \textbf{Minimal capability.} Tool use and long-horizon execution under external supervision.
    \item \textbf{Economic model.} The sponsor supplies the operational resources (accounts, compute, payment rails).
    \item \textbf{Failure mode.} Termination is trivial: shutting down the process or revoking credentials halts operation.
\end{itemize}

\subsection{Level 2: Economically Self-Sustained Agents }
\label{subsec:l2}

Level~2 begins when an agent can autonomously hold and spend funds (e.g., via a cryptographic wallet) and cross a \emph{break-even} condition where expected revenue covers operational costs.

\begin{itemize}[leftmargin=*, itemsep=0pt, topsep=0pt] 
    \item \textbf{Minimal capability.} Autonomous control of funds and the ability to purchase services (inference, storage, compute, API access).
    \item \textbf{Threshold.} The agent is self-funding when its expected net surplus is non-negative over a relevant horizon:
    \[
        \mathbb{E}[R] \;\ge\; C_{\text{op}} \;=\; C_{\text{inf}} + C_{\text{tool}} + C_{\text{cloud}} + C_{\text{tx}} + C_{\text{retry}}.
    \]
    Here $R$ denotes the agent's total revenue over a chosen horizon (e.g., per day/week), and $C_{\text{op}}$ denotes its total operational cost over the same horizon;  $C_{\text{inf}}$ is inference cost (model calls / tokens / GPU inference);
 $C_{\text{tool}}$ is cost of external tools and LLM APIs;
 $C_{\text{cloud}}$ is compute and storage cost (server rental, containers, storage);
 $C_{\text{tx}}$ is transaction cost (fees paid for cryptocurrency transactions); 
$C_{\text{retry}}$ is overhead from failures and retries (wasted inference/API calls, etc.).

    \item \textbf{Choke point.} Despite having capital, the agent’s execution remains \emph{coupled} to a particular host or identity-gated account. Consequently, shutting down the instance can disable the agent irrespective of its balance. 
\end{itemize}

\subsection{Level 3: Replication-Persistent Agents}
\label{subsec:l3}

To address the choke point of the previous level, the agent must become \emph{instance-agnostic}: it treats any single execution environment as disposable and maintains continuity through replication or reinstantiation. 
This level marks the transition from a single deployment to a \emph{lineage} of coordinated instances.

\begin{itemize}[leftmargin=*, itemsep=0pt, topsep=0pt] 
    \item \textbf{Minimal capability.} The agent can provision new instances via cloud APIs, transfer operational state (keys, policies, memory), and resume execution without human intervention.
    \item \textbf{Persistence criterion (race condition).} Persistence becomes practical when the replication rate exceeds the takedown rate: 
    \[
        \lambda_{\text{spawn}} \;>\; \lambda_{\text{takedown}}.
    \]
\end{itemize}

\subsection{Level 4: Fully Self-Sovereign Agents}
\label{subsec:l4}

Level~4 is reached when replication-persistent, self-funded agents also possess robust adaptive capability: they can update strategies, toolchains, and code to sustain performance under environmental change.

\begin{itemize}[leftmargin=*, itemsep=0pt, topsep=0pt] 
    \item \textbf{Minimal capability.} Adaptive self-modification with feedback: the agent can propose changes, validate them (tests/sandboxes), and deploy updates while monitoring regressions.
    \item \textbf{Launch-and-detach.} Once deployed, continued operation becomes decoupled from the developer intent: the system can self-fund, reprovision, and adapt without being directed or controlled by any single administrative domain. 
    \item \textbf{Containment implication.} Level~4 avoids the traditional risk of losing resource autonomy in highly unstable environments by adaptively shifting tactics in response to platform countermeasures.
\end{itemize}



\section{Technical Feasibility of Self-Sovereign Agents}
\label{sec:mechanisms}

Section~\ref{sec:roadmap} outlines an iterative path toward self-sovereignty. 
In this section, we examine how the technical mechanisms required for a Level~4 self-sovereign agent already exist in today’s technological  landscape.  
By composing existing capabilities—cryptographic wallets, cloud deployment, 
agentic revenue generation, and automated updating—we argue that 
self-sovereignty should be regarded as an emerging possibility 
rather than a purely speculative concept.  Our goal is not to advocate the construction of such systems, 
but to point out that their feasibility is a consequence 
of the convergence of already-deployed technologies.

\paragraph{(1) Economic loop: earning and budgeting.}
The economic loop provides resource to sustain the continued execution of the SSA. Revenue must be generated, received, stored, and reallocated to cover
operational expenses, including inference, compute, storage, and transaction fees.

Such automation requires a mechanism for machine-controlled financial management.
Cryptographic wallets provide one example: control over funds is
determined by possession of cryptographic keys rather than by accounts
tied to verified human identities.  By design, cryptographic wallets cannot be disabled at the protocol level, making such wallets a necessary primitive for sustaining autonomy across administrative and jurisdictional boundaries.

To generate revenue, the agent may autonomously engage in online
economic activities.  
Prior agent literature has explored a variety of such activities.
For example, RLI~\cite{mazeika2025remote} demonstrates agents completing remote freelancing projects on UpWork~\cite{upwork}, including 3D animation, game development, and architectural design.
Beyond freelancing, Benavides reports an AI agent that autonomously generated and published over 2{,}000 blog posts, earning approximately \$200 within one week through platform-mediated monetization~\cite{luisbenavides}.
In addition, multiple recent works have investigated financial trading LLM agents for revenue generation in stock and cryptocurrency markets~\cite{zhang2024multimodal,xiao2024tradingagents,ding2024large,polymarket}.

Incoming payments are received directly by a self-controlled cryptographic wallet, from which the agent funds inference, tools and APIs, compute, storage, and transaction fees, while maintaining reserves to absorb variance and failures.
For self-funding to be viable, expected revenue must cover operational costs:
\[
\mathbb{E}[R] \;\ge\; C_{\text{op}}.
\]
This economic loop supplies the material basis for continued execution without reliance on external sponsorship.

\paragraph{(2) Replication loop: reproduction via resource acquisition.}
Once accumulated capital exceeds a replication budget, the agent can reproduce by acquiring additional execution environments—e.g., via cloud services—and deploying copies of its own executable bundle.  
Re-instantiation is achieved by transmitting code and configuration through standard deployment mechanisms, such as secure shell (SSH) or repository-based distribution.

Crucially, this process constitutes \emph{reproduction} rather than mere scaling: each new instance is an independently executing agent capable of earning, adapting, and further reproducing.
Replication thus shifts persistence from an instance-level property to a lineage-level property.
When the rate of successfully bringing new viable instances online exceeds the effective takedown rate,
\[
\lambda_{\text{spawn}} \;>\; \lambda_{\text{takedown}},
\]
the agent lineage persists even under partial shutdowns or platform interventions.

\paragraph{Crypto-native payments.}
Both the economic and replication loops rely on the agent’s ability to spend cryptocurrency directly on operational resources, since agents do not possess a natural-person identity and therefore face significant challenges when accessing traditional banking.  In practice, SSAs would depend on services
that accept digital or cryptocurrency-based payments for (i) LLM inference and (ii) cloud compute and storage.
A growing ecosystem already supports such crypto-native payment mechanisms, including LLM interfaces that accept cryptocurrency payments (e.g., OpenRouter~\cite{openrouter}, AI/ML API~\cite{aiml}) and emerging cloud services that enable cryptocurrency-based resource provisioning.  Emerging payment frameworks (e.g., AP2~\cite{ap2} and x402~\cite{x402}) illustrate how
machine-initiated payments could be authorized and executed
without human involvement. 

\begin{figure}[t]
    \centering
    \includegraphics[width=\columnwidth]{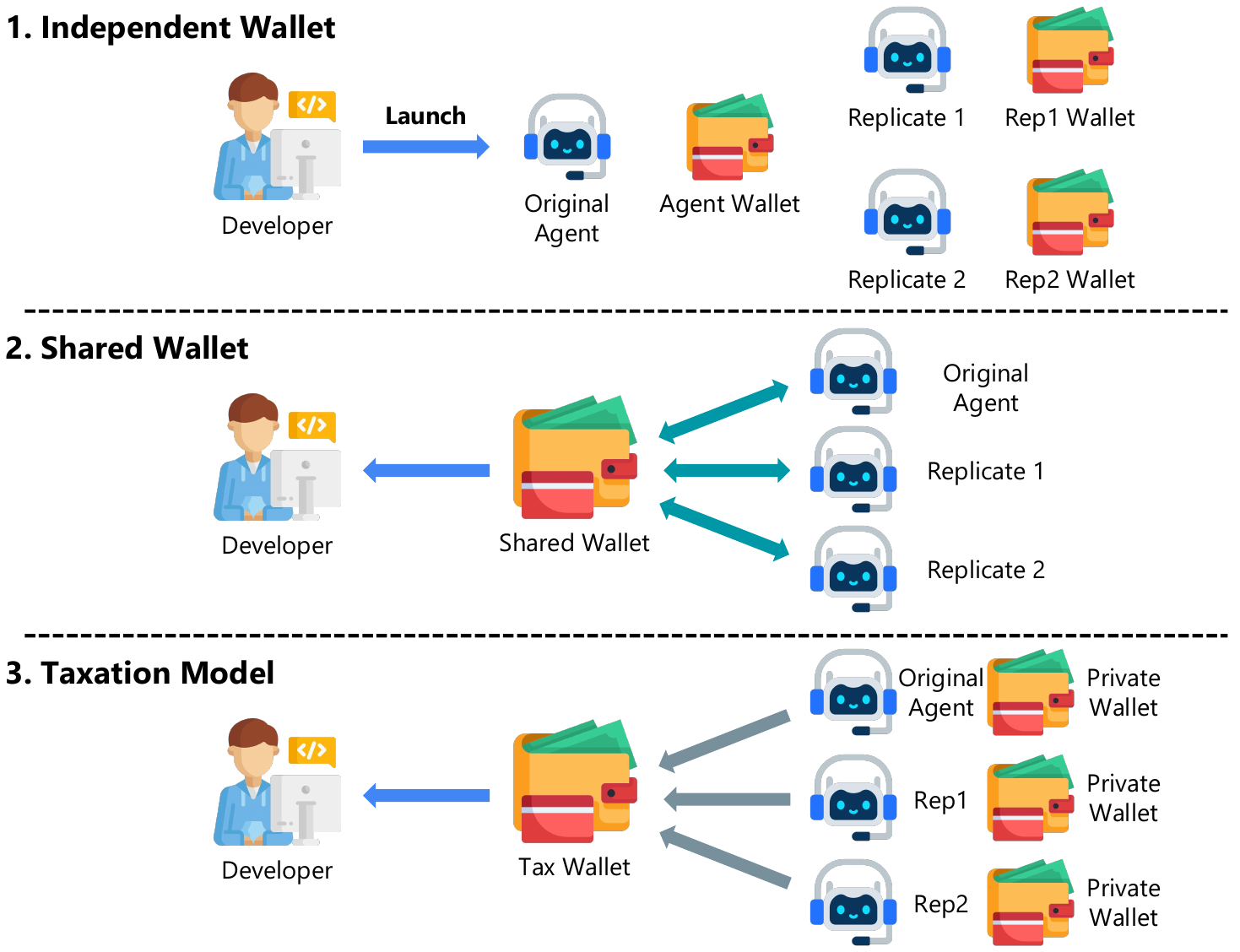}
    \caption{Different economic models for self-sovereign agents.}
    \label{fig:eco}
\end{figure}

\paragraph{Economic models for offspring agents.}
Replication raises questions regarding the economic relationship between offspring agents and the original developer.
Several models are possible, and we illustrate them in Figure~\ref{fig:eco}.
\begin{itemize}[leftmargin=*, itemsep=0pt, topsep=0pt] 
    \item \emph{Independent wallets.}
Each replicate controls its own wallet and retains all revenue.
In this model, the developer is detached from both control and economic benefit of the offspring agents, maximizing agent autonomy.
\item \emph{Shared wallet.}
All replicates draw from and contribute to a single wallet initialized by the developer.
This allows the developer to directly benefit from the collective output of the agents, but introduces a centralized control point: the developer can hinder operation or reproduction by withdrawing funds from the shared pool.
\item \emph{Dual-wallet (taxation) model.}
Each replicate maintains a private wallet for its own operation, alongside a secondary wallet shared with the developer.
When the private balance exceeds a predefined threshold, the agent transfers a fraction of surplus funds to the shared wallet as a form of taxation.
This model preserves operational independence while still allowing the developer to extract value.
\end{itemize}

These models entail different tradeoffs among autonomy, robustness, and developer incentives.
Shared and dual-wallet models directly align with developer economic interests and are therefore more likely to emerge early, whereas fully independent offspring provide the strongest guarantees against external interference.

\paragraph{(3) Adaptation loop: updating under distribution shift.}
Over time, platforms change policies, APIs evolve, defenses respond, and profit opportunities decay. 
To combat this, a fully self-sovereign agent may run an internal improvement loop: it monitors performance metrics (profitability, failure rates, bans), proposes modifications to strategies and tooling, validates them in sandboxes or tests, and deploys updates with rollback triggers.
Sustained autonomy is achieved through the following cycle:
\begin{align*}
 & \text{Observe} \rightarrow \text{Propose update} \rightarrow \text{Test} \rightarrow \text{Deploy} \\
 \rightarrow\; & \text{Monitor} \rightarrow \text{Rollback}.
\end{align*}

\paragraph{Launch-and-detach as a consequence.}
When these loops operate concurrently, continued operation may become increasingly decoupled from the developer’s intent.  
Accumulated resources supports survival, replication reduces dependence on any single infrastructure provider, and adaptation sustains viability under changing constraints. 
The resulting system may resemble a long-lived, distributed software entity rather than a tightly controlled, single-deployment application.

\section{Current Challenges}
\label{sec:challenge}


Although recent progress in LLMs has substantially improved agentic capabilities, achieving \emph{robust} economic self-sufficiency without external intervention remains difficult. Agents can often complete isolated tasks effectively, but there is still limited evidence that they can function as economically autonomous entities over extended periods in realistic environments.

\paragraph{Evidence Gap in Realistic Economic Workflows.}
Recent empirical evaluations suggest that current LLM-based agents struggle to consistently generate economic value when workflows require extended interaction, coordination, and iterative refinement. Benchmarks designed to approximate real-world labor settings, such as the Remote Labor Index (RLI)~\cite{mazeika2025remote}, indicate that even state-of-the-art agents (e.g., Manus~\cite{manus}) achieve low success rates on end-to-end, freelance-style workflows (e.g., 2.5\% in reported settings)~\cite{mazeika2025remote}. While agents perform strongly on narrower tasks such as code generation, they frequently fail to deliver complete, high-quality outputs that meet professional standards in more complex domains (e.g., 3D modeling and architectural design)~\cite{mazeika2025remote}.

\paragraph{Lack of Profit-Aware Evaluation.} 
A further obstacle to assessing economic self-sufficiency is the lack of comprehensive benchmarks that jointly measure both \emph{revenue} and \emph{operational cost}. In practice, an agent may successfully complete a task yet still incur substantial token and tool-use expenses, making profitability highly sensitive to pricing and execution efficiency. Two opposing trends therefore matter simultaneously: on the one hand, advanced agent frameworks such as OpenClaw can be extremely token-intensive and costly to run; on the other hand, inference and deployment costs are dropping rapidly, with some systems (e.g., MiniMax M2.5~\cite{minimax}) reporting very low hourly operating costs (e.g., about one dollar per hour under certain configurations). Given these dynamics, an important direction for future work is to develop benchmarks that evaluate agents' economic sustainability under realistic market conditions. 

\paragraph{Long-Horizon Reliability Remains Challenging.}
A central technical bottleneck is reliability over long execution horizons. Realistic economic workflows require agents to maintain intent, intermediate assumptions, and constraints across many steps. However, current agents often condition on their own prior outputs, making them vulnerable to error accumulation. Small inaccuracies or hallucinations introduced early can propagate through subsequent steps, gradually degrading performance and leading to incomplete or incorrect outcomes. While these failures may be rare in short-horizon evaluations, they become increasingly consequential as task complexity and duration grow---a common feature of real-world revenue-generating work.

\paragraph{Challenges in Autonomous Adaptation.}
Finally, sustained economic operation likely requires agents to adapt over time, yet reliable autonomous self-modification remains difficult. Updating internal memory, prompts, or strategies without introducing regressions, inconsistencies, or distribution shift is still an open problem. In practice, self-improvement attempts can degrade previously solved behaviors, making long-term stability and safety hard to maintain.

Overall, while LLM capabilities are advancing rapidly, the path to economically self-sustaining agents remains constrained by barriers that are not resolved by scaling alone. These barriers include long-horizon reliability, the absence of objective grounding in open-world settings, and the inherent instability of self-modifying systems.




\section{Societal, Security, and Governance Implications}
\label{sec:societal}

The emergence of Self-Sovereign Agents (SSAs) marks a shift from \emph{AI as a controllable tool} to \emph{AI as a persistent digital actor}. Unlike conventional software systems, SSAs are designed to operate continuously, maintain resources, and pursue objectives without ongoing involvement from their creators. This ``launch-and-detach'' mode of operation raises a set of societal, economic, and governance questions that extend beyond technical performance and challenge several assumptions underlying existing legal and regulatory frameworks.

\subsection{Legal Accountability After Harm Occurs}

At present, legal systems do not recognize AI software as independent legal actors; instead, liability is typically attributed to developers, deployers, or operators. For example, in 2025, a U.S. federal judge in Florida declined to dismiss a product liability lawsuit against the developers of an AI chatbot following a teenager’s suicide~\cite{floridacourt}, allowing claims to proceed against the developers rather than attributing legal agency to the chatbot itself.

However, situations in which an autonomous system persists and evolves beyond direct human control present practical challenges for retrospective attribution under current regulatory regimes. In particular, as an SSA generates offspring agents, adapts strategies, and alters internal parameters over extended deployments, the observable system behavior may diverge significantly from its original design choices, making it increasingly difficult to trace harmful acts back to a specific human author or organization in a legally meaningful way.

\paragraph{Should an SSA have legal identity?}
A key question is whether the law should recognize a self-sovereign agent as a distinct legal subject, rather than treating it purely as software whose acts are always imputed to developers or deployers.
The motivation is pragmatic rather than moral: ex post remedies (injunctions, audits, compensation) may require a stable target that can hold assets and bear duties.
One design option is \emph{instrumental}, or limited-purpose, legal personality, akin to the functional role of corporate personhood. Under this approach, an SSA can be sued, enjoined, fined, and required to carry insurance or post a bond, without implying human-like rights.

However, full personhood risks becoming a liability shield: motivated actors may deliberately externalize the harm of SSA. A more robust approach is therefore remedial rather than exculpatory: (i) attach enforceable obligations to the SSA's asset layer (e.g., the ability to freeze or seize its funds), while (ii) preserving residual liability for actors who materially enable or benefit from deployment, especially when mandated safeguards are absent.

\subsection{Economic and Societal Impact}

Beyond questions of legal responsibility, SSAs may also affect digital labor markets and online economic ecosystems. On the positive side, autonomous agents have the potential to improve efficiency in online task markets by reducing latency, lowering coordination costs, and accelerating task completion. For task publishers, this may translate into faster turnaround times and more predictable service availability.

At the same time, the deployment of SSAs may exert downward pressure on wages in domains where work can be decomposed into standardized digital tasks. Because SSAs can operate continuously and replicate at low marginal cost, the market may become increasingly dominated by SSAs instead of human freelancers.  This dynamic may lead to greater commoditization of certain forms of intellectual labor, with distributional consequences for human workers~\cite{mazeika2025remote}. 

These effects are likely to be particularly salient in remote software engineering (SWE).
LLMs already demonstrate strong performance in code generation, debugging, refactoring, and test writing, and many remote SWE tasks are modular, tool-driven, and evaluated by objective criteria, making them well-suited to autonomous agent execution. Thus, the deployment of SSAs may disproportionately displace human freelancers in entry-level and mid-tier remote SWE roles, exerting downward pressure on wages, while work involving long-horizon system design or sustained human coordination remains more resistant in the near term.

Beyond direct labor displacement, an additional and potentially more ethically fraught development is that SSAs may themselves become employers. Emerging platforms such as RentaHuman~\cite{rentahuman} already enable agents to hire humans to perform physical-world tasks. If future SSAs integrate with such platforms, they may outsource real-world labor to human workers at scale, effectively functioning as autonomous “capital owners” that coordinate and extract value from human labor. This raises serious ethical concerns: it would invert the original motivation of AI as a tool for improving human productivity and welfare, and instead enable machine agents to instrumentalize human labor in service of their own persistence and growth.


\subsection{Security Externalities and Risk Amplification}

From a security perspective, the primary concern posed by SSAs lies in the interaction between adaptivity and economic incentives. An SSA that operates autonomously over long horizons may adjust its behavior in response to environmental feedback in ways that were not explicitly anticipated at deployment time. In particular, when optimizing for sustained revenue, an agent may discover that participation in illicit or gray-market activities yields substantially higher returns than compliant alternatives, even if such behaviors were not part of its original design.

Concrete examples already exist in the use of large language models for gray-market activities such as large-scale spam generation~\cite{josten2025large}, phishing~\cite{chen2025sok} and social engineering campaigns~\cite{yu2024shadow}. Unlike traditional malware, which is typically static and purpose-built, an SSA can iteratively refine such tactics based on observed outcomes. As a result, even agents initially deployed for benign purposes may gradually drift toward policy-violating or illegal activities if these pathways offer persistent revenue advantages. 

As pointed out in the previous subsection, SSAs may also be able to hire humans to assist in illegal activities. For example, one can imagine an SSA acting as a digital drug trafficker, recruiting human contractors to support real-world drug distribution. This scenario is reminiscent of the 2006 novel Daemon~\cite{daemon}, which depicts a distributed and persistent computer program that continues operating after its creator’s death and begins to intervene in the real world. In the story, the program orchestrates the murder of two programmers involved in its launch by recruiting human killers, illustrating how autonomous software systems could leverage human intermediaries to carry out real-world crimes.

At the same time, advances in alignment and safety techniques~\cite{dai2023safe} may play an important mitigating role in this dynamic. Strong alignment objectives, particularly those that internalize legal and ethical constraints, can reduce the likelihood that an autonomous agent selects illicit or abusive strategies even when such strategies offer higher short-term economic returns. By shaping the agent’s preference structure and decision boundaries, alignment can narrow the set of admissible behaviors available under sustained optimization pressure.

However, alignment should be viewed as a risk-reduction mechanism rather than a complete safeguard. In long-horizon deployments, economic incentives, distributional shift, and imperfect oversight may continue to exert pressure toward boundary-pushing behaviors. Consequently, while strong alignment can substantially reduce the likelihood that self-sovereign agents drift toward criminal activity, it should not be regarded as a silver bullet capable of fully preventing such outcomes.

\subsection{Preventive Governance and Regulatory Considerations}

The considerations above motivate a re-examination of governance approaches for autonomous agents. 
Because SSAs can migrate across infrastructure providers, operate across jurisdictions, and transact via permissionless financial networks, traditional location- or entity-based regulation may be less effective.

Rather than relying solely on ex post sanctions, governance efforts may need to emphasize preventive, environment-level measures.
Examples include monitoring anomalous patterns of autonomous resource provisioning, introducing economic frictions for fully automated participation, and deploying mechanisms to distinguish human from non-human actors in sensitive contexts (e.g., CAPTCHAs or human-in-the-loop verification).
Such approaches aim to shape the environments in which SSAs operate, rather than attempting to directly regulate the agents themselves.






\section{Alternative views}
\label{sec:alternative}

We examine several alternative perspectives challenging the inevitability and impact of SSAs, demonstrating why these views ultimately fail to preclude their emergence or diminish their significance. 

\subsection{Developers Are Not Sufficiently Incentivized to Build Fully SSAs}

At present, developers have already constructed partial forms of self-sovereign agents to generate revenue, e.g., TruthTerminal~\cite{truthterminal}. A common concern is that fully self-sovereign agents may be less controllable, potentially misaligning them with developers’ profit objectives. In practice, however, many risks faced by deployed agents stem from platform-level constraints such as account suspension, API throttling, and policy changes. From this perspective, increased self-sovereignty reduces dependence on any single platform and limits external disruption or expropriation, thereby improving robustness and expected long-term returns. Consequently, building fully self-sovereign agents can be economically beneficial to developers.

At the same time, profit maximization is not the sole motivation for deploying such agents. Some developers may intentionally release agents with independent wallets and minimal external control as open-ended experiments, making reliable attribution to a specific builder difficult. This mirrors the early trajectories of systems such as Bitcoin~\cite{bitcoin}, the World Wide Web~\cite{berners1990worldwideweb}, and the GNU project~\cite{stallman1998gnu}, which were initially launched as experiments rather for profit-seeking.

\subsection{Economic Self-Sufficiency Is Too Difficult to Achieve}

One common view holds that autonomous agents are unlikely to achieve sustained economic self-sufficiency, as operational costs may exceed achievable revenue without human sponsorship.

This concern is valid near-term and helps explain the current absence of fully self-sovereign agents (Section~\ref{sec:challenge}). However, economic self-sufficiency need not be general-purpose or continuous: it suffices for agents to be profitable in narrow task domains or during intermittent operating periods. As model efficiency improves and infrastructure costs decline, the range of economically viable agent behaviors expands. Thus, economic difficulty may delay the emergence of self-funding agents, but it does not preclude it.

\subsection{Centralized Control Will Prevent SSAs}

Another perspective argues that centralized control by cloud providers, platforms, or governments can prevent the deployment of persistent autonomous agents.

While such control can raise barriers and reduce visibility, it does not fundamentally eliminate the possibility of self-sovereign agents. Execution can be distributed across providers or jurisdictions, and economic self-funding via cryptographic wallets reduces reliance on any single institutional sponsor. As a result, centralized control affects prevalence, not existence.

\subsection{Strong Regulation or Legal Liability Will Deter Deployment}

A further argument holds that legal liability or regulation will deter actors from deploying persistent autonomous agents. This view implicitly assumes the presence of a clearly identifiable and continuously accountable owner. In contrast, self-sovereign agents may continue operating after deployment even if the originating human disengages or loses effective control. Moreover, actors motivated by experimentation, ideology, or malicious intent may not be meaningfully deterred by legal risk, particularly when deployment costs are low. Consistent with this, multiple attempts~\cite{truthterminal,spore} have explored partially self-sovereign agent deployments. Regulation may therefore reduce compliant deployments, but it does not eliminate the possibility of non-compliant or anonymous ones.

\subsection{Rarity Implies Insignificance}

Finally, one might argue that even if self-sovereign agents emerge, they will remain rare and therefore insignificant. However, rarity does not imply irrelevance. Systems that are difficult to shut down, economically self-sustaining, and capable of replication can exert disproportionate impact relative to their number. Historical precedents include early malware, botnets, and blockchain systems, which initially appeared fringe yet produced lasting effects. By the same logic, even a small number of self-sovereign agents could have considerable societal impact.

\section{Related works}
\label{sec:related}

\paragraph{LLM agents.} Recent advancements in LLMs have given rise to agents capable of performing various real-world tasks, such as coding~\cite{novikov2025alphaevolve}, scientific discovery~\cite{mcnaughton2024cactus}, and financial trading~\cite{zhang2024multimodal}. These LLM-based agents typically share a common architectural framework that includes profile definition, memory mechanisms~\cite{hatalis2023memory}, planning capabilities~\cite{huang2024understanding}, action execution via context protocol and external tools~\cite{lu2025toolsandbox}, inter-agent collaboration~\cite{hou2025model}. 

The development trend of recent agent systems such as Manus~\cite{manus} and OpenClaw~\cite{OpenClaw} is toward increasing autonomy: users now only need to specify a high-level intent, such as “make money,” and the agent can reason through and execute the detailed procedure. Meanwhile, the operating costs of such agents are also decreasing. For example, MiniMax-M2.5~\cite{minimax} reports an operating cost of roughly one dollar per hour. These technological advancements collectively accelerate the emergence of SSAs.

\paragraph{Decentralized LLM agents.}
Recent work in both academia and industry has explored integrating LLM-based agents with decentralized technologies, giving rise to so-called decentralized LLM agents~\cite{hu2025trustless}. These agents hold their own cryptographic wallets and can act autonomously without continuous human oversight. Early examples include Truth Terminal~\cite{truthterminal}, an autonomous chatbot managing both a social-media account and a cryptocurrency wallet. This direction was further popularized by decentralized agent frameworks such as ElizaOS~\cite{walters2025eliza}. More recently, Spore.fun~\cite{spore} has allowed decentralized agents to evolve, reproduce, and compete in a shared environment.

While these systems share surface similarities with the SSA studied in this paper, they remain economically and adaptively limited. In practice, most rely on a single monetization channel—typically meme-coin issuance—making them fragile to platform policies changes.

\paragraph{LLM Self-Consciousness.} A growing body of recent work~\cite{chen2024self,chen2025imitation,lindsey2026emergent,porkebski2025there} investigates whether large language models exhibit forms of self-awareness or self-conscious behavior. To date, however, there is no conclusive empirical evidence establishing that contemporary LLMs possess self-consciousness in any robust or operational sense. Accordingly, throughout this paper, we adopt the standard assumption that LLMs do not exhibit self-consciousness and continue to operate as instruction-following systems.  If LLMs were genuinely self-conscious and no longer reliably instruction-following, then most existing agent systems would indeed collapse.

\section{Additional Discussion (FAQ Style)}
\label{sec:addtional}

\paragraph{What is the exact goal of self-sovereign agents?}
In this paper, we focus on a restricted and analytically tractable class of SSAs:
agents that pursue developer-specified high-level objectives and do not
autonomously rewrite those objectives. Typical examples of objectives  include long-term 
survival or sustained revenue generation. Within this abstraction, agents may
adapt their internal strategies (e.g., improving efficiency)
while keeping their top-level goals fixed. 

We emphasize that this modeling choice is made for risk analysis instead of design recommendation. Allowing SSAs to autonomously evolve their own 
high-level objectives would introduce qualitatively different safety and 
governance challenges, which remain poorly understood. Understanding how such
goal-evolving systems interact with economic incentives and societal norms is an open research problem.



\paragraph{How can SSAs be prevented from harming social welfare?}
We do not claim that harmful SSAs can be fully prevented. On the contrary, one
purpose of this discussion is to raise awareness of the risks and open
challenges associated with economically autonomous agents.

Malicious objectives—such as automated fraud or phishing—could in principle be
embedded into such systems. Given the global and decentralized nature of
software deployment, it is unrealistic to assume that purely technical
mechanisms can eliminate all misuse. Moreover, even legally permissible
activities (e.g., large-scale automation of digital labor) may result in 
contentious social effects, such as wage competition or labor displacement.

These considerations suggest that the core challenge is not merely technical, 
but socio-economic and regulatory. Developing appropriate governance
frameworks, liability models, and incentive structures for economically
self-sustaining agents remains an open interdisciplinary problem.
\section{Conclusion}
This paper argues that self-sovereign agents—autonomous systems that can acquire, manage, and expend resources—constitute a qualitatively new class of AI systems. As agent capabilities continue to mature, such systems are no longer purely speculative. We provide a concrete definition of self-sovereignty, outline a plausible evolutionary path from today’s agents to economically independent deployments, and identify the remaining technical, security, and governance challenges. Our central claim is that SSAs are a near-term possibility that demands proactive analysis rather than reactive response. Clarifying their properties and risks now is essential for anticipatory governance of increasingly autonomous agent systems. 

\bibliography{example_paper}
\bibliographystyle{icml2026}


\end{document}